# Detection of Ultra-Trace Heavy metals in Aerosols with pg/m³ Sensitivity Using Filament-Induced Fluorescence Spectroscopy

*Yuezheng Wang[1,2,#], Lu Sun[1,2,#], Zhiwenqi An[1,2], Jiayun Xue[1,2], Zhixuan An[1,2], Nan Zhang[1,2], Lie Lin[1,3], Weiwei Liu[1,2,*]*

[1]Institute of Modern Optics, Eye Institute, Nankai University, Tianjin 300350, China

[2]Tianjin Key Laboratory of Micro-scale Optical Information Science and Technology, Tianjin 300350, China

[3]Tianjin Key Laboratory of Optoelectronic Sensor and Sensing Network Technology, Tianjin 300350, China

**E-mail:** liuweiwei@nankai.edu.cn

Heavy metal pollution, particularly in the form of airborne aerosols such as lead (Pb), cadmium (Cd), mercury (Hg), and cobalt (Co), poses serious health and environmental risks, necessitating highly sensitive remote detection techniques. In this study, Filament-Induced Fluorescence Spectroscopy (FIFS) was employed to detect ultra-trace concentrations of heavy metal aerosols with high sensitivity and stability. By systematically optimizing the balance between filament length and detection distance, the optimal detection distance under the current experimental conditions was determined. With a detection distance of 10 m, this work achieved a minimum detectable concentration of 0.47 pg/m³ for Pb and an extrapolated limit of detection (LOD) of 0.3 pg/m³, with excellent signal stability (RSD < 7%) over a concentration range from 0.47 pg/m³ to 0.47 µg/m³. Additionally, Cd, Hg, and Co aerosols were also successfully detected under the same conditions, with detection limits of 2 pg/m³, 0.25 pg/m³, and 3 pg/m³, respectively, further confirming the versatility of FIFS in detecting diverse heavy metals. Theoretical predictions suggest that increasing laser power could further enhance the detection capability. These results highlight the ultra-sensitive remote detection capability of FIFS for heavy metal aerosol detection and provide valuable insights for optimizing system parameters to enhance its application performance in environmental monitoring.

## 1. Introduction

In recent years, heavy metal pollution in the atmosphere, particularly in the form of aerosols containing toxic elements such as lead (Pb), cadmium (Cd), mercury (Hg), and cobalt (Co), has drawn significant attention due to its severe health threats.[1,2] These heavy metals are among the most hazardous pollutants, with Pb, for instance, being linked to irreversible health issues such as allergies, anemia, cognitive impairment, and damage to the kidneys, nervous system, reproductive system, liver, and brain.[3-8] Exposure to heavy metal aerosols originates from a wide range of anthropogenic activities, including coal combustion, mining, chemical production, metal refining, electroplating, battery manufacturing, waste incineration, and vehicular emissions.[9,10] Heavy metal aerosols are particularly concerning due to their ability to remain suspended in the air for long periods, causing inhalation exposure and long-distance transport, making remote sensing technologies crucial for monitoring the distribution of heavy metal aerosols.

Various heavy metals, including Pb, Cd, Hg, Co, are increasing regulatory attention due to their toxicity, persistence, and potential for bioaccumulation. Among them, Pb has been the most extensively studied, and serves as a representative case for understanding the dangers of long-term low-level exposure. The United States Environmental Protection Agency (EPA) has set an annual average limit of 0.15 µg/m³ for Pb in total suspended particles (TSP),[11] and the Codex Alimentarius Commission specifies a maximum Pb concentration of approximately 1 ng/m³ in infant food.[12] Studies show that even trace levels of airborne Pb can lead to significant health effects in vulnerable populations such as children.[13] For instance, continuous exposure to a Pb aerosol concentration of just 1 pg/m³ over 2-3 years may increase a child's blood lead level to the Centers for Disease Control and Prevention (CDC) reference threshold of 5 µg/dL. Moreover, approximately 90-95% of the lead absorbed by the human body from exposure to lead-containing aerosols is stored in the bones, where it has a long half-life of 20-30 years, leading to sustained health effects.[14] Considering that Cd, Hg, and Co exhibit similar toxicological effects and bioaccumulation characteristics, achieving pg/m³-level or lower detection sensitivity is essential for the early identification of heavy metal aerosols and accurate environmental risk assessment.

Currently, analytical methods for detecting heavy metal aerosols include Laser-Induced Breakdown Spectroscopy (LIBS),[15] Atomic Absorption Spectrometry (AAS),[16] X-ray Fluorescence (XRF),[17] Surface Enhanced Raman Spectroscopy (SERS),[18] Colorimetric method,[19] Enzyme-Linked Immunosorbent Assay (ELISA),[20] Anodic Stripping Voltammetry (ASV),[21] Fast-Scan Cyclic Voltammetry (FSCV),[22,23] Differential Pulse Voltammetry (DPV)[24] and Square Wave Voltammetry (SWV)[25]. Among these, LIBS, AAS, XRF, SERS, and Colorimetric methods are classified as optical detection methods, ELISA is categorized as a biochemical method, and SWV, FSCV, DPV, and ASV are classified as an electrochemical detection method. While these techniques offer high measurement accuracy, lack remote sensing capabilities, making it difficult to meet the demand for real-time, precise monitoring of heavy metal aerosols under complex environmental conditions.[26] For example, LIBS uses high-energy nanosecond laser pulses to generate plasma through rapid ablation and ionization of material, requiring short focal lengths to maintain laser intensity, and it is difficult to generate plasma in suspended particles, making it unsuitable for aerosol remote sensing.[27] Therefore, the importance of remote sensing technologies in heavy metal aerosol detection has become increasingly apparent, as pollutants such as Pb, Cd, Hg, and Co exhibit high atmospheric mobility and long-range transport potential.

In response to these limitations and to further enhance detection accuracy, new improved methods have been developed, including aerosol trapping techniques (e.g., optical trapping, acoustic focusing, and static field concentration),[28,29] dual-pulse configurations,[30] and laser-induced filamentation.[31] Among these, Filament-Induced Fluorescence Spectroscopy (FIFS) leverages femtosecond (fs) laser pulses to form stable, high-intensity laser filaments over long distances. This enables efficient plasma generation in aerosol particles, overcoming the diffraction limitations of nanosecond lasers and enhancing laser-sample coupling. FIFS works by utilizing the dynamic balance between Kerr self-focusing and plasma defocusing caused by multiphoton or tunneling ionization when powerful femtosecond laser pulses propagate through the air.[32] This balance maintains the laser intensity inside the filament core at approximately $5 \times 10^{13}$ W/cm², which is sufficient to dissociate or ionize most materials and produce characteristic fluorescence. These characteristics make FIFS a highly promising technique for real-time, high-sensitivity elemental analysis. By enabling remote detection of heavy metal aerosols, it allows for continuous monitoring of large or hard-to-reach areas, such as industrial zones, highly polluted urban areas, or remote ecological sensitive regions, without the need for direct sampling.

In this study, we applied Filament-Induced Fluorescence Spectroscopy for the sensitive and reliable detection of trace lead aerosols. The detection limit of FIFS depends primarily on the fluorescence intensity generated by the filament in aerosols and the efficiency of fluorescence collection by the detector. To analyze the factors affecting fluorescence behavior, we systematically investigated the gain from amplified spontaneous emission (ASE) and the loss caused by Mie scattering during filament-aerosol interactions. By optimizing the balance between filament length and detection distance, we established optimal experimental conditions for remote detection. Under a 10-meter detection distance, FIFS achieved pg/m³-level sensitivity for all tested heavy metal aerosols, with the minimum detectable concentration for Pb reaching 0.47 pg/m³ and an extrapolated LOD of 0.3 pg/m³, further confirming the broad applicability of FIFS in trace heavy metal aerosol monitoring. To the best of our knowledge, compared to other optical detection methods, electrochemical methods, and biochemical methods, the FIFS we use has significant advantages in detection limit and remote sensing capability.

## 2. Materials and methods

### 2.1. Preparation of Aerosol Samples

All materials and chemicals used in this study were of analytical grade and were used without further purification. The primary standard solutions for dilution included single-element standard solutions of Pb (1 mg/L in 2% nitric acid), Cd (1 mg/L), Hg (1 mg/L), and Co (1 mg/L), all purchased from Bolinda (Shenzhen, China). Ultrapure water (18.2 MΩ·cm, Shanghai Swan Biotechnology Co., Ltd.) was used for all dilutions.

Initially, standard solutions of Pb, Cd, Hg, and Co were stepwise diluted to generate solutions with concentrations ranging from 0.1 ng/L to 1 mg/L. These diluted solutions were nebulized into aerosols using an aerosol generator (HRH-WAG3, Beijing HuiRongHe Technology Co., Ltd.), producing particles with a mass median diameter of 0.1~3 µm, with the aerosol generator and associated components shown in Figure 1. The resulting aerosol was directed into a glass tube (3 cm diameter, 1 m length), where an adjustable air pump ensured a stable flow during testing.

Using the extinction method, the resulting heavy metal aerosols were calibrated to air concentrations of 0.047 pg/m³ to 0.47 µg/m³ for Pb, 0.6 pg/m³ to 0.6 µg/m³ for Cd, 0.056 pg/m³ to 0.56 µg/m³ for Hg, and 0.9 pg/m³ to 0.9 µg/m³ for Co. All aerosol generation and testing were performed in a fume hood to safeguard experimenters from heavy metal toxicity.

### 2.2. Filament-Induced Fluorescence Spectroscopy Experimental Setup

The FIFS system used in experiment consisted of three main components: a laser source, filamentation optical components, and a detection apparatus, whose overall architecture is depicted in the central and lower sections of Figure 1.

The Magma200 laser (Amplitude) operates at a repetition rate of 100 Hz with a pulse energy adjustable from 0 to 200 mJ. It delivers ultrashort pulses with a full width at half maximum (FWHM) of 496.9 fs and a peak power exceeding 400 GW/cm², and a wavelength of 1030 nm.

To induce filamentation, the laser beam is first reflected by Mirror 1 and then focused by a telescope system comprising a concave lens ($f = -15$ cm; $R = 2.54$ cm) and a concave mirror ($f = 200$ cm; $R = 20.7$ cm), as shown in Figure 1. By adjusting the relative distance between these components, the starting point of the filament can be precisely controlled, ranging from 2 to 40 meters. The filament interacts with heavy metal aerosols, generating plasma that emits characteristic fluorescence.

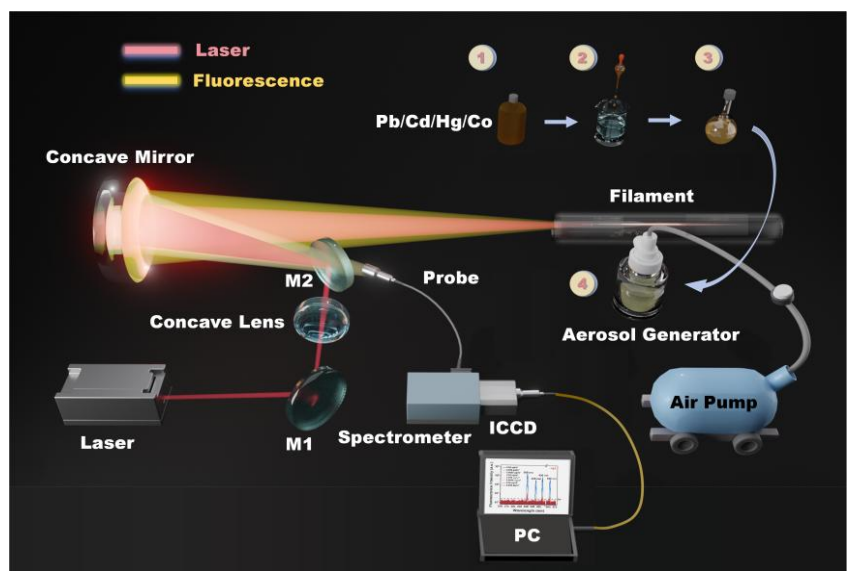

**Figure 1.** Schematic diagram of the Filament-Induced Fluorescence Spectroscopy (FIFS) system for heavy metal aerosol detection. The upper-right section illustrates the preparation process for heavy metal aerosols using an aerosol generator. The main components of the FIFS system, shown in the central and lower sections, include: M1 (plane mirror), M2 (dichroic mirror), concave lens, concave mirror, spectrometer, ICCD, PC and air pump.

The detection apparatus comprises five components: a concave mirror, an optical fiber, a spectrometer, an Intensified Charge-Coupled Device (ICCD), and a computer. The backward fluorescence signals generated by the interaction between the filament and aerosols are collected and focused by the concave mirror, then transmitted through an optical fiber probe positioned behind Mirror 2, a dichroic mirror (LBTEK Co., Ltd., China) with over 95% reflectance at 1030 nm and over 95% transmittance in the 350-550 nm range. The optical fiber transmits the fluorescence to a grating spectrometer (Omni-λ300, Zolix) covering a wavelength range of 200~1100 nm with a resolution of 0.2 nm. The spectral data is recorded by an Istar-sCMOS camera and stored on a computer.

To ensure high signal-to-noise ratio spectra, the ICCD parameters were set to a gate delay of 220 ns, a gate width of 1000 ns, and an exposure time of 30 seconds. The spatial position of the optical fiber's receiving end was meticulously optimized using a manual 3D translation stage to maximize fluorescence collection efficiency.

## 3. Results and Discussion

### 3.1. Fluorescence Spectrum and Limit of Detection

We performed fluorescence spectral detection of heavy metal aerosols at a detection distance of 10 m. The measurement results, shown in Figure 2, clearly display the prominent fluorescence emission lines of metal atoms. When the filament interacts with the heavy metal aerosols, the metal ions ($Pb^{2+}$, $Cd^{2+}$, $Hg^{2+}$, $Co^{2+}$) in the aerosols are subjected to the intense electric field of the femtosecond laser, reducing them to their corresponding metal atoms (Pb, Cd, Hg, Co).[33] In the high-energy density region of the filament, these metal atoms are excited, causing their electrons to transit from the ground state to higher excited states, such as from $6s^2\ 6p^2$ to $6s^2\ 6p\ 7s$ for Pb, from $4d^{10}\ 5s\ 5p$ to $4d^{10}\ 5s\ 5d$ for Cd, from $5d^{10}\ 6s\ 6p$ to $5d^{10}\ 6s\ 6d$ for Hg, and from $3p^6\ 3d^8\ 4s$ to $3p^6\ 3d^8\ 4p$ for Co.[34] As the excited electrons return to lower energy levels, fluorescence photons are emitted, forming the fluorescence emission spectra for Pb, Cd, Hg, and Co, each with characteristic peaks.

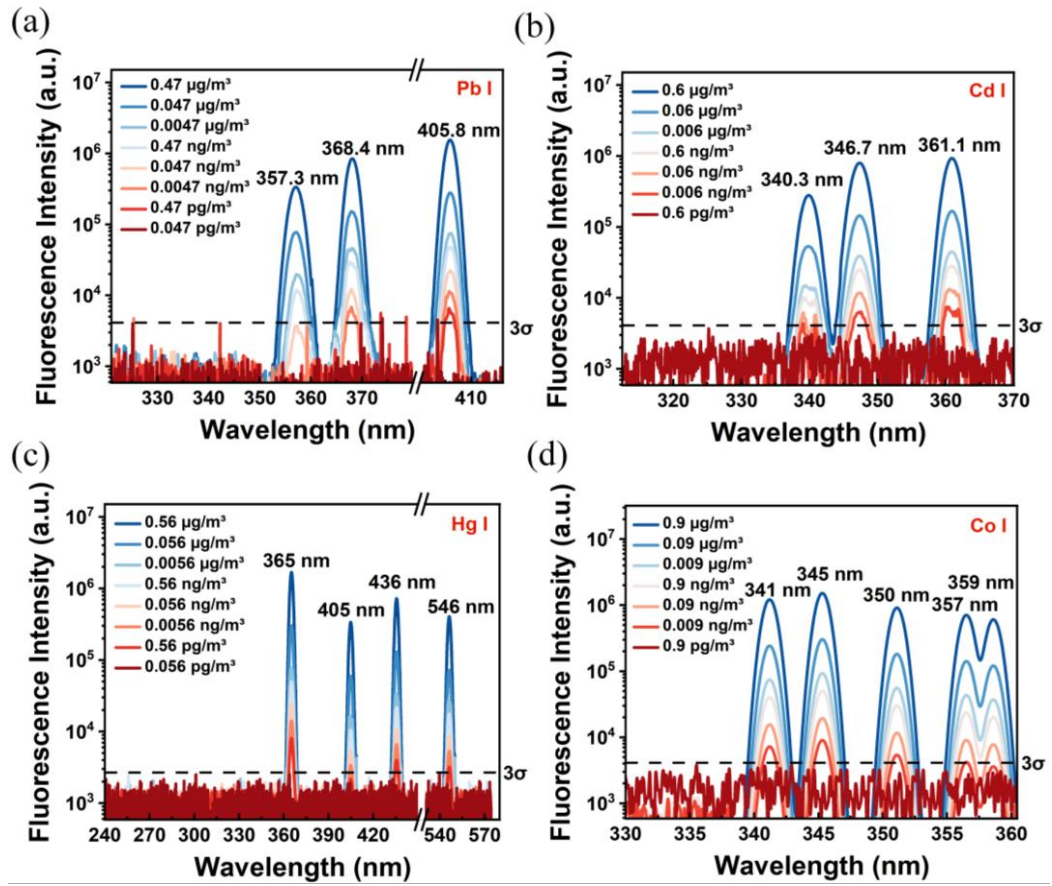

**Figure 2.** Fluorescence spectra of heavy metal aerosols at varying concentrations. (a) Pb: 0.047 pg/m³ to 0.47 µg/m³, with peaks at 357.3 nm, 368.4 nm, and 405.8 nm; (b) Cd: 0.6 pg/m³ to 0.6 µg/m³, with peaks at 340.3 nm, 346.7 nm, and 361.1 nm; (c) Hg: 0.056 pg/m³ to 0.56 µg/m³, with emissions at 365 nm, 405 nm, 436 nm, and 546 nm; (d) Co: 0.9 pg/m³ to 0.9 µg/m³, with peaks at 341 nm, 345 nm, 350 nm, 357 nm and 359 nm. The 3σ level represents three times the standard deviation of the background noise.

For the Pb, Cd, Hg, and Co aerosols, prominent fluorescence peaks are observed at 357.3 nm, 368.4 nm, and 405.8 nm for Pb, at 340.3 nm, 346.7 nm, and 361.1 nm for Cd, at 365 nm, 405 nm, 436 nm, and 546 nm for Hg, and at 341 nm, 345 nm, 350 nm, 357 nm, and 359 nm for Co, as shown in Figure 2. According to the National Institute of Standards and Technology (NIST) database, the fluorescence wavelengths of Pb, Cd, Hg, and Co match well with our experimental measurements. As the concentration of heavy metal aerosols increases, the intensity of these spectral peaks increases correspondingly, as shown by the color distribution in the spectra. Among these peaks, the fluorescence at 405.8 nm for Pb I, 361.1 nm for Cd I, 365 nm for Hg I, and 345 nm for Co I exhibit the highest relative intensities. These wavelengths were therefore selected for concentration calibration and subsequent measurements.

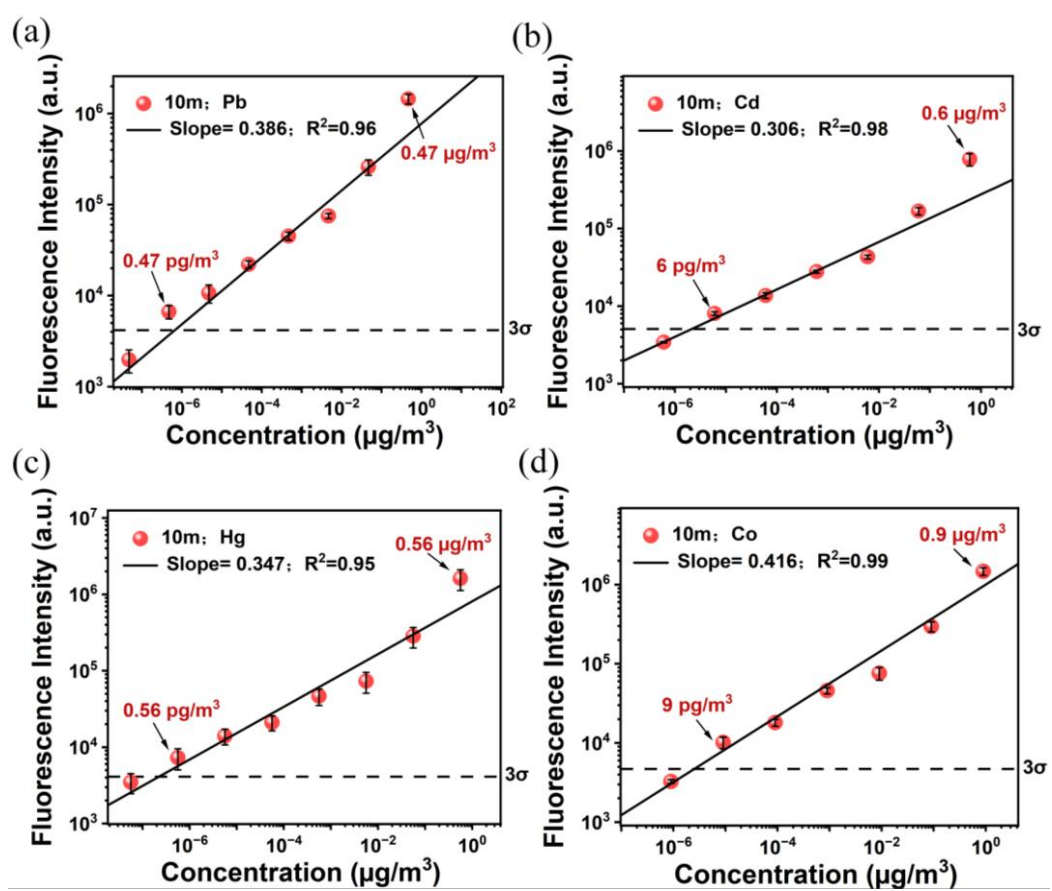

**Figure 3.** Calibration curves for heavy metal aerosols at 10 m: (a) Pb: 0.47 pg/m³ to 0.47 µg/m³, with a slope of 0.386 and R² = 0.96; (b) Cd: 6 pg/m³ to 0.6 µg/m³, with a slope of 0.306 and R² = 0.98; (c) Hg: 0.56 pg/m³ to 0.56 µg/m³, with a slope of 0.347 and R² = 0.95; (d) Co: 9 pg/m³ to 0.9 µg/m³, with a slope of 0.416 and R² = 0.99. The 3σ level represents three times the standard deviation of the background noise.

To quantitatively determine the detection limit, the integrated fluorescence intensity of the fluorescence spectral lines at the respective strongest wavelengths for Pb, Cd, Hg, and Co was extracted and fitted against the metal aerosol concentration, generating the calibration curve shown in Figure 3. In the experiment, the directly measured minimum concentrations of Pb, Cd, Hg, and Co aerosols were 0.47 pg/m³, 6 pg/m³, 0.56 pg/m³, and 9 pg/m³, respectively. The calibration curves show a linear relationship across concentration ranges of 0.047 pg/m³ to 0.47 µg/m³ for Pb, 0.6 pg/m³ to 0.6 µg/m³ for Cd, 0.056 pg/m³ to 0.56 µg/m³ for Hg, and 0.9 pg/m³ to 0.9 µg/m³ for Co, with R² values of 0.96, 0.98, 0.95, and 0.99, respectively. According to the recommendations of the International Union of Pure and Applied Chemistry (IUPAC), the limit of detection (LOD) can be calculated using the calibration curve obtained in Figure 3 with the following expression:

$$\text{LOD} = \frac{3\sigma}{S} \quad (1)$$

Here, $\sigma$ represents the standard deviation of the signal intensity from blank samples, reflecting the system's noise level. $S$ denotes the slope of the fitted calibration curve between concentration and signal intensity. Based on these parameters, the LOD for Pb in heavy metal aerosols was calculated to be 0.3 pg/m³ from Figure 3a, 2 pg/m³ for Cd from Figure 3b, 0.25 pg/m³ for Hg from Figure 3b, and 3 pg/m³ for Co from Figure 3d. The experimental measurements are in good agreement with the values calculated from the calibration curves, demonstrating the reliability of the detection system.

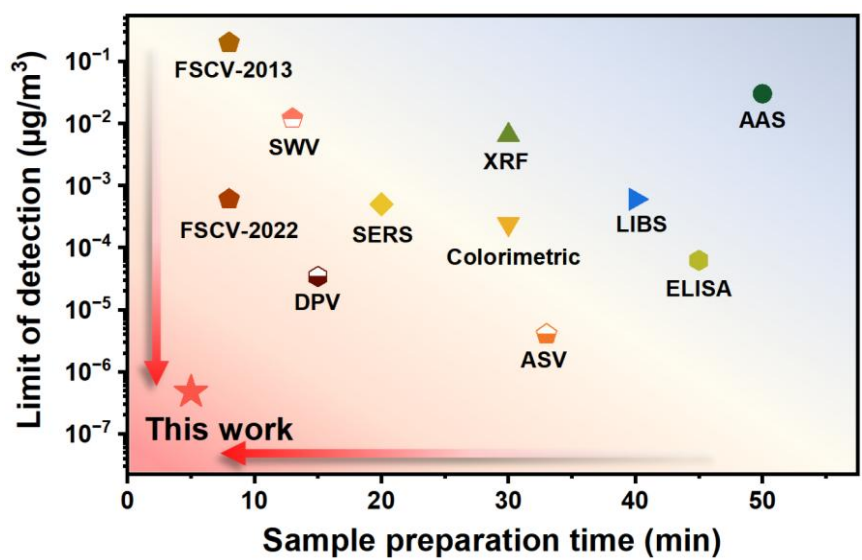

**Figure 4.** Comparison of detection limits and sample preparation times for various heavy metal aerosol detection methods. The chart includes Atomic Absorption Spectrometry (AAS), Laser-Induced Breakdown Spectroscopy (LIBS), X-ray Fluorescence (XRF), Surface Enhanced Raman Spectroscopy (SERS), Colorimetric methods, Enzyme-Linked Immunosorbent Assay (ELISA), Square Wave Voltammetry (SWV), Fast-Scan Cyclic Voltammetry (FSCV) Differential Pulse Voltammetry (DPV), Anodic Stripping Voltammetry (ASV), and the method used in this work (FIFS).

Based on literature research,[15-25] the detection limits and sample preparation times for mainstream heavy metal aerosol detection methods vary, as shown in Figure 4: AAS (30 ng/m³, 50 min), LIBS (0.6 ng/m³, 40 min), ELISA (0.062 ng/m³, 45 min), XRF (6.5 ng/m³, 30 min), Colorimetric (0.24 ng/m³, 30 min), ASV (0.004 ng/m³, 33 min), SERS (0.05 ng/m³, 20 min),

FSCV (200 ng/m³, 8 min, 2013), FSCV (0.6 ng/m³, 8 min, 2022), SWV (12 ng/m³, 13 min) and DPV (0.034 ng/m³, 15 min). While these techniques provide high measurement accuracy, they generally lack remote sensing capabilities, which makes real-time, precise monitoring of heavy metal aerosols in complex environments challenging. In contrast, FIFS not only offers remote sensing capabilities but also demonstrates superior detection limits and sample preparation times, making it particularly well-suited for such applications.

## 3.2. The Stability of Filament-Induced Fluorescence Spectroscopy

Signal stability is a critical parameter for any detection technology. In this study, the signal stability is quantified using the relative standard deviation (RSD), which is calculated as follows:

$$\mathrm{RSD} = \frac{SD}{\bar{x}} \times 100\% \quad ; \quad SD = \sqrt{\frac{\sum_{i=1}^{n}(x_i - \bar{x})^2}{n-1}} \tag{2}$$

Here, $SD$ represents the standard deviation, and $\bar{x}$ denotes the mean value. To assess signal stability, twelve repeated measurements were conducted on Pb, Cd, Hg, and Co aerosols, with concentration ranges of 0.47 pg/m³ to 0.47 µg/m³ for Pb, 6 pg/m³ to 0.6 µg/m³ for Cd, 0.56 pg/m³ to 0.56 µg/m³ for Hg, and 9 pg/m³ to 0.9 µg/m³ for Co, using FIFS.

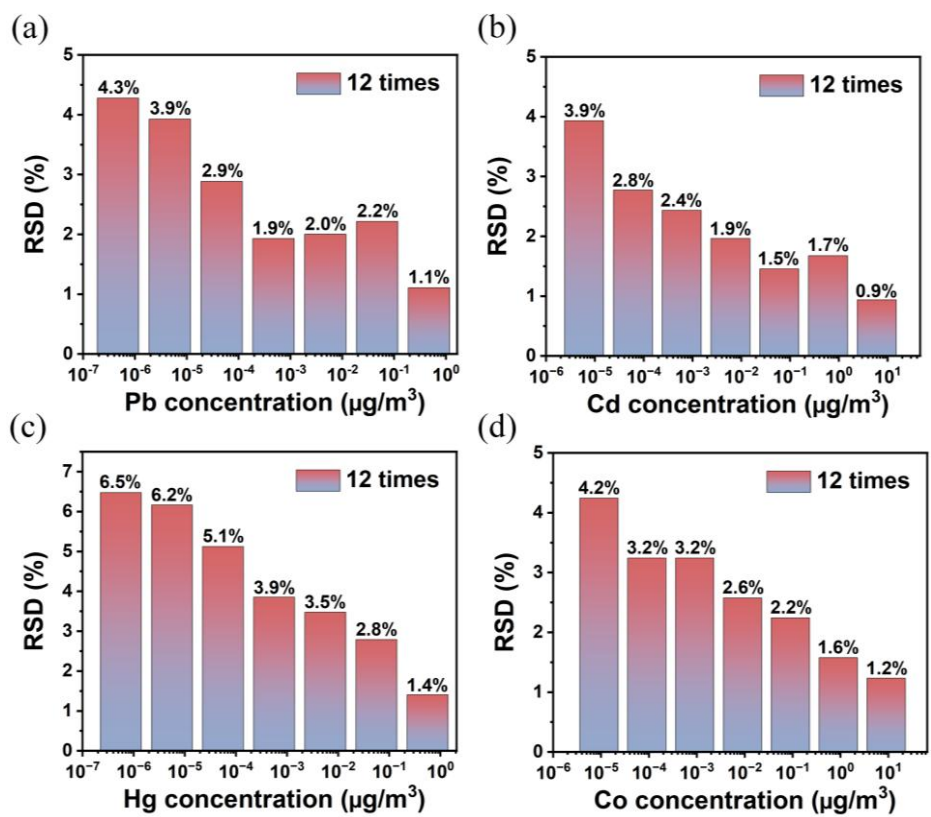

**Figure 4.** Relative standard deviation (RSD) of fluorescence intensity for heavy metal aerosols at various concentrations over twelve repeated measurements: (a) Pb; (b) Cd; (c) Hg; (d) Co.

As shown in Figure 4a, the measurements of Pb in heavy metal aerosols reveal that the RSD decreases from 4.3% at a concentration of 0.47 pg/m³ to 1.1% at 0.47 µg/m³. At lower Pb concentrations, the fluorescence signal exhibits greater fluctuation due to the influence of background noise. However, as the Pb concentration increases, the fluorescence signal stability improves. This trend underscores the system's robust performance across a wide dynamic range. Figures 4b to 4d show similar trends for Cd, Hg, and Co aerosols, where the fluorescence signal stability improved as the concentration increased. For Cd, the RSD of twelve measurements decreased from 3.9 % at 6 pg/m³ to 0.9 % at 0.6 µg/m³. For Hg, the RSD decreased from 6.5 % at 0.56 pg/m³ to 1.4 % at 0.56 µg/m³. For Co, the RSD decreased from 4.2 % at 9 pg/m³ to 1.2 % at 0.9 µg/m³. Overall, the RSD for all heavy metal aerosols remained below 7% across a wide concentration range, confirming the reliability of the FIFS system for detecting heavy metal aerosols.

### 3.3. Mechanism of the Fluorescence

FIFS utilizes high-intensity femtosecond laser pulses to generate filaments in air, which interact with the target aerosol, inducing fluorescence through strong-field ionization, dissociation, and excitation processes. Taking Pb in heavy metal aerosols as an example, the fluorescence generation mechanism of Pb is discussed. As shown in Figure 5, Pb has an atomic number of 82, with an electron configuration of [Xe] $4f^{14}\ 5d^{10}\ 6s^2\ 6p^2$, and the outermost electrons are in the $6p$ orbital. Upon exposure to the high-intensity filament, the laser energy interacts with these outer electrons, causing the electrons to transition from the ground state $6p^2\ ^3P_0$ (0 eV) to the excited state $6p^1\ 7s\ ^3P_1^o$ (4.37 eV). Subsequently, the excited electrons radiatively relax to a lower energy level, $6p^2\ ^3P_2$ (1.32 eV), emitting a photon with a wavelength of 405.8 nm. This transition has a relatively high transition probability ($A_{ki} \approx 9\times10^7\ s^{-1}$), allowing the electrons to more rapidly transition from the excited state to the lower energy level, thereby releasing more photons and enhancing the fluorescence signal intensity. The resulting fluorescence spectra provide characteristic information of Pb atoms, while the fluorescence intensity is directly proportional to the concentration of Pb in the aerosol.

Due to the ultrashort duration of femtosecond pulses, free electrons cannot efficiently absorb energy from the laser field via inverse bremsstrahlung, resulting in a plasma with relatively low temperature and electron density. This effectively suppresses non-specific collisional emission and continuum background, thereby enhancing the signal-to-noise ratio (SNR), making FIFS particularly suitable for trace heavy metal aerosols detection. Moreover, owing to the intensity clamping effect of the filament, each point along its path acts a dynamic "focal region". As the filament length increases, the accumulated fluorescence signals significantly enhance the overall intensity. As illustrated in Figure 5, previous studies have revealed that this enhancement exhibits an exponential dependence on filament length, primarily attributed to the mechanism of amplified spontaneous emission (ASE).[35] The presence of ASE not only reinforces the fluorescence signal along the filament path but also substantially improves the detection sensitivity of the FIFS technique.

In the following sections, we continue to use Pb in heavy metal aerosols as an example and systematically examine the underlying physical mechanisms that contribute to the enhanced performance of FIFS in the detection of trace heavy metal aerosols.

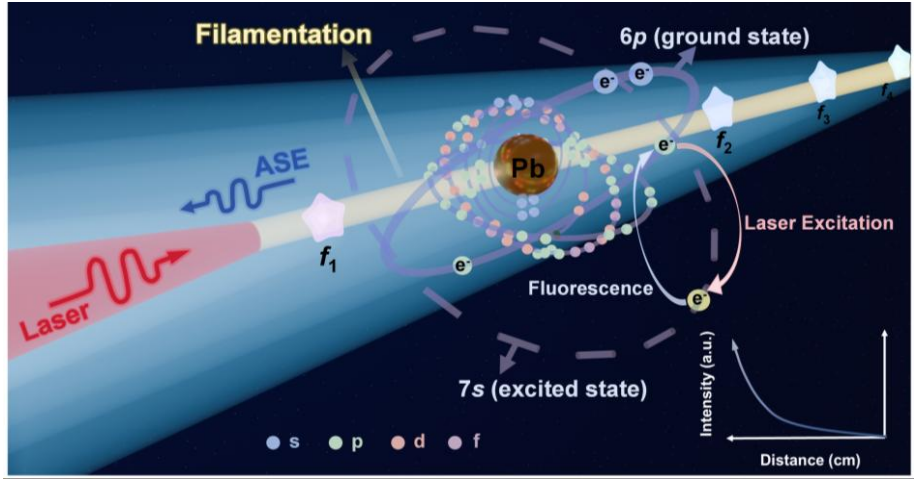

**Figure 5.** Schematic of femtosecond filament-induced fluorescence and enhancement via amplified spontaneous emission (ASE). The filament excites Pb atoms, causing electron transitions from the ground state (6$p$) to the excited state (7$s$). Upon relaxation to a lower energy level, fluorescence is emitted. Each point along the filament acts as a dynamic 'focal region' ($f_1$, $f_2$, $f_3$, ...), with fluorescence intensity increasing exponentially with filament length, primarily due to ASE.

The gain and loss of backscattered fluorescence signals induced by filament in aerosols have been widely studied. The gain originates from ASE, and the backscattered fluorescence intensity $I$ can be expressed as follows:[36]

$$I = \frac{P_s}{g}\left(e^{gL} - 1\right) \quad (3)$$

where $P_s$ represents the spontaneous emission power per unit length, $g$ is the optical gain coefficient, and $L$ is the filament length. The optical gain coefficient g is defined as:[37]

$$g = N \cdot \sigma \quad (4)$$

Where $N$ is the number density of excited Pb atoms, and σ is the transition radiation cross-section. At lower Pb concentrations, the number density of excited lead atoms decreases, resulting in a smaller $g$. At the same time, when the filament length is limited, the fluorescence intensity gradually transitions from an exponential relationship to an approximately linear relationship with the filament length. The loss of fluorescence signals mainly originates from the influence of Mie scattering by aerosol particles before the fluorescence signals reach the detector. The propagation loss of fluorescence is represented by the extinction coefficient $K_2$, which characterizes the attenuation of fluorescence per unit length caused by particles in the medium. Its expression is given as:[38,39]

$$I = I_0 e^{-K_2 L} \tag{5}$$

Here, $I_0$ is the fluorescence initial intensity, and $L$ still represents the filament length. Figure 6a presents the fitting results where both ASE gain and Mie scattering loss are incorporated into the model. The model exhibits excellent agreement with the experimental data, achieving R² values exceeding 0.99 across all Pb aerosol concentrations.

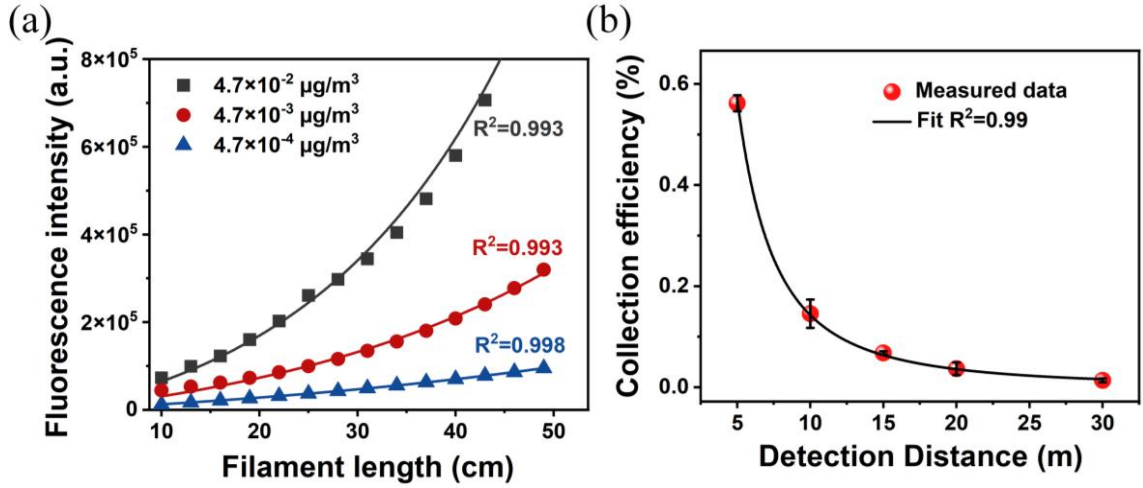

**Figure 6.** (a) Fluorescence intensity as a function of filament length for different Pb aerosol concentrations; (b) Collection efficiency as a function of detection distance, following an inverse-square relationship.

Although the fluorescence intensity increases with the filament length, extending the filament length in the FIFS system is typically achieved by increasing the incident laser energy or the effective focal length of the telescope system. The impacts of these two methods will be discussed separately below.

Increasing the effective focal length also extends the detection distance, leading to a reduction in the fluorescence intensity received by the detector due to the inverse-square law. As shown in Figure 6b, we measured the variation in fluorescence intensity with detection distance for a fixed filament length, confirming this inverse-square relationship.

To achieve the strongest fluorescence signal under the current experimental conditions, a trade-off must be made between filament length and detection distance. At low Pb concentrations, the fluorescence intensity exhibits an approximately linear relationship with filament length, while the intensity decreases with detection distance following an inverse-square law. By constructing a mathematical function that incorporates these two factors, the optimal point can be determined by finding the maximum through differentiation. Specifically, the total fluorescence intensity received by the detector can be expressed as:

$$I_{total} = (aL + b) \cdot \frac{c}{D^2} \quad (6)$$

$$L = D - \frac{0.367 L_{DF}}{\sqrt{\left[(P_{in}/P_{cr})^{1/2} - 0.852\right]^2 - 0.0219}} \quad (7)$$

$I_{total}$ represents the total fluorescence intensity received by the detector. The term a$L$+b describes fluorescence excitation, where $L$ is the filament length, a is a proportional coefficient, and b is the background intensity when the filament length is zero. The term c/$D^2$ accounts for collection efficiency, where $D$ is the detection distance and c is a proportional coefficient related to the fluorescence collection efficiency. This function integrates both the linear enhancement of fluorescence with filament length and the inverse-square reduction in signal with detection distance. Equation (7) expresses the relationship between filament length and focal length (which can also be regarded as the detection distance).[40] Specifically, $L_{DF}$ represents the Rayleigh length, describing the distance over which a beam remains focused before it starts to diverge, $P_{in}$ is the input power, which depends on the laser pulse energy and the repetition rate. $P_{cr}$ is the critical power, representing the power threshold at which the beam begins to filament.

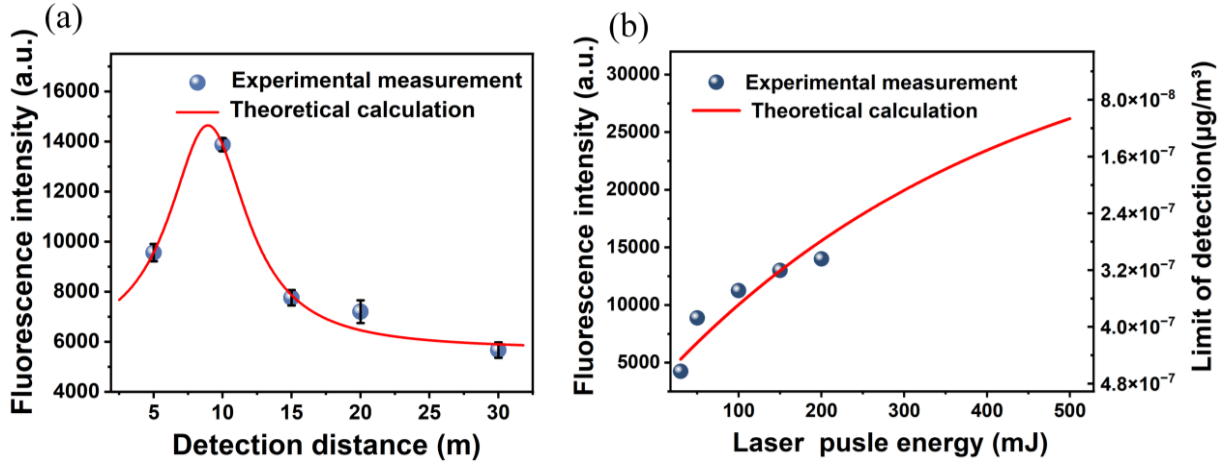

**Figure 7.** (a) Fluorescence intensity as a function of detection distance for Pb aerosols at a concentration of 47pg/m³. The experimental measurements (blue spheres) show a peak at 10 m, consistent with the theoretical calculation; (b) Fluorescence intensity as a function of laser pulse energy, with theoretical calculations (red curve). The right axis shows the corresponding limit of detection.

By substituting Equation (7) into Equation (6), a function is obtained with the detection distance $D$ as the independent variable and the total fluorescence intensity received by the detector as the dependent variable. Under the current experimental conditions, by analyzing the first and second derivatives of this function, the optimal filament length is determined to be 57 cm, corresponding to a detection distance of 10 m. To validate the effectiveness of this function, we measured the fluorescence intensity of Pb aerosols at a concentration of 47 pg/m³ across different detection distances. As shown by the blue spheres in Figure 7a, the fluorescence intensity first increases and then decreases with distance, reaching a maximum near 10 m. This experimental result is consistent with the theoretical prediction, confirming the validity of the model.

The limit of detection for aerosols is also influenced by variations in laser pulse energy. As shown in Figure 7b, in our experiment, we measured the Pb fluorescence intensity at a concentration of 47 pg/m³ across different laser pulse energies, including 30 mJ, 50 mJ, 100 mJ, 150 mJ, and 200 mJ. For the theoretical validation, the filament length can be calculated using Equation (7), and then, using Equation (3) and Equation (5), the fluorescence intensity received by the detector and the corresponding detection limit at different laser pulse energies can be determined. Assuming the use of a more powerful ultrafast laser source, as the single-pulse laser energy increases, the fluorescence intensity also increases, leading to further improvements in the limit of detection. For example, when the laser pulse energy increases from 200 mJ to 500 mJ, the theoretical LOD improves by an order of magnitude.

## 4. Conclusion

In this study, Filament-Induced Fluorescence Spectroscopy (FIFS) was applied for the sensitive and stable detection of heavy metal aerosols. By optimizing the trade-off between filament length and detection distance, the FIFS system achieved its best performance, with a detection limit (LOD) for Pb in heavy metals of 0.3 pg/m³ and a minimum detectable concentration of 0.47 pg/m³ at 10 m away. The relative standard deviation (RSD) remained below 7% across a broad concentration range, demonstrating the system's high stability. In addition to Pb, the system also successfully detected Cd, Hg, and Co aerosols, all with experimentally measurable detection thresholds at the pg/m³ level, further confirming its versatility. Future advancements in higher-energy lasers could further elongate the sensing distance and enhance the LOD by improving fluorescence intensity. These findings underline the robustness, high sensitivity, and remote sensing capability of FIFS, providing a quantitative foundation for system optimization and highlighting its potential for trace heavy metal aerosol detection in environmental monitoring and atmospheric remote sensing applications.

## Acknowledgements

This work was supported by the National Natural Science Foundation of China (W2412044, 12204251,12304381), Russian Science Foundation (25-49-00154) and Fundamental Research Funds for the Central Universities (63243166)

## Author Contributions

Weiwei Liu conceived the research plan. Yuezheng Wang and Weiwei Liu designed the research. Jiayun Xue conducted the theoretical calculations related to filament. Yuezheng Wang, Zhixuan An, and Zhiwenqi An were responsible for the experimental system setup and detection limit measurements. Yuezheng Wang analyzed the data. Yuezheng Wang, Lu Sun, and Weiwei Liu discussed the results. Nan Zhang and Lie Lin supervised the work. Yuezheng Wang and Lu Sun co-wrote the manuscript. All authors participated in the discussion of the results and provided feedback on the manuscript.

## Conflict of Interest

The authors declare no conflict of interest.

## Data Availability

The data supporting this study's findings are available from the corresponding author upon reasonable request.